\documentclass[preprint,aps]{revtex4}
\usepackage{mathrsfs}
\usepackage{graphicx}
\usepackage{multirow}
\usepackage{makecell}
\usepackage{booktabs}
\usepackage{color}
\usepackage{bm}

\begin{document}

\title{Giant and Reversible Electronic Structure Evolution in a Magnetic Topological Material EuCd$_2$As$_2$}

\author{Yang Wang$^{1,2\sharp}$, Cong Li$^{1,2\sharp}$, Taimin Miao$^{1,2\sharp}$, Shuai Zhang$^{1,2\sharp}$, Yong Li$^{1,2\sharp}$, Liqin Zhou$^{1,2}$, Meng Yang$^{1,2,3}$, Chaohui Yin$^{1,2}$, Yongqing Cai$^{1,2}$, Chunyao Song$^{1,2}$, Hailan Luo$^{1,2}$, Hao Chen$^{1,2}$, Hanqing Mao$^{1}$, Lin Zhao$^{1,2,4}$, Hanbin Deng$^{1,2}$, Yingkai Sun$^{1,2}$, Changjiang Zhu$^{1,2}$, Fengfeng Zhang$^{5}$, Feng Yang$^{5}$, Zhimin Wang$^{5}$, Shenjin Zhang$^{5}$, Qinjun Peng$^{5}$, Shuheng Pan$^{1,2,4,6}$, Youguo Shi$^{1,2,3}$, Hongming Weng$^{1,2,4}$, Tao Xiang$^{1,2,4,7}$, Zuyan Xu$^{5}$  and X. J. Zhou$^{1,2,4,7*}$
}

\affiliation{
\\$^{1}$Beijing National Laboratory for Condensed Matter Physics, Institute of Physics, Chinese Academy of Sciences, Beijing 100190, China
\\$^{2}$University of Chinese Academy of Sciences, Beijing 100049, China
\\$^{3}$Center of Materials Science and Optoelectronics Engineering, University of Chinese Academy of Sciences, Beijing 100049, China
\\$^{4}$Songshan Lake Materials Laboratory, Dongguan 523808, China
\\$^{5}$Technical Institute of Physics and Chemistry, Chinese Academy of Sciences, Beijing 100190, China
\\$^{6}$CAS Center for Excellence in Topological Quantum Computation, University of Chinese Academy of Sciences, Beijing 100190, China
\\$^{7}$Beijing Academy of Quantum Information Sciences, Beijing 100193, China
\\$^{\sharp}$These people contributed equally to the present work.
\\$^{*}$Corresponding authors:  XJZhou@iphy.ac.cn
}
\date{August 26, 2022}


\maketitle

%%Abstract

{\bf The electronic structure and the physical properties of quantum materials can be significantly altered by charge carrier doping and magnetic state transition. Here we report a discovery of a giant and reversible electronic structure evolution with doping in a magnetic topological material. By performing high-resolution angle-resolved photoemission measurements on EuCd$_2$As$_2$, we found that a huge amount of hole doping can be introduced into the sample surface due to surface absorption. The electronic structure exhibits a dramatic change with the hole doping which can not be described by a rigid band shift. Prominent band splitting is observed at high doping which corresponds to a doping-induced magnetic transition at low temperature (below $\sim$15\,K) from an antiferromagnetic state to a ferromagnetic state. These results have established a detailed electronic phase diagram of EuCd$_2$As$_2$ where the electronic structure and the magnetic structure change systematically and dramatically with the doping level. They further suggest that the transport, magnetic and topological properties of EuCd$_2$As$_2$ can be greatly modified by doping. These work will stimulate further investigations to explore for new phenomena and properties in doping this magnetic topological material.
}

%%Introduction
The electronic structure and physical properties of a material can be manipulated by varying the crystal structure, magnetic structure and the concentration of charge carriers. The doping of charge carriers can cause significant change of electronic structures and resultant physical properties like insulator-metal transition and emergence of superconductivity\cite{MImadaRMP,PLeeRMP}. EuCd$_2$As$_2$ is a novel magnetic topological material\cite{GYHua_PRB2018,LLWang_PRB,CWNiu_PRB,JZMa_AM,JZMa_SA}. A number of topological states including antiferromagnetic topological insulators\cite{JZMa_AM}, a triple-point magnetic topological semimetal\cite{GYHua_PRB2018} and a quantum anomalous Hall insulator\cite{CWNiu_PRB} can be produced in EuCd$_2$As$_2$ by manipulating its magnetic structures combined with rotational or inversion symmetry breaking. In particular, when EuCd$_2$As$_2$ is in the inter-layer antiferromagnetic state, it is expected to host only one pair of Dirac points at the Fermi level\cite{GYHua_PRB2018}. When the ferromagnetism is induced along the c axis, it can generate a single pair of Weyl points that is an ideal platform to study the unique properties of Weyl semimetals\cite{LLWang_PRB,JZMa_SA,Soh_PRB2019}. EuCd$_2$As$_2$ has therefore become a fertile playground for studying the interplay between magnetism and topology. The magnetic properties of EuCd$_2$As$_2$ originate from the Eu layers and there have been extensive studies on manipulating its magnetic structure by temperature\cite{Artmann_Chem,Schellenberg_Chem,MCRahn_PRB,JZMa_SA,JZMa_AM,Jo_PRB,YXu_PRB,YWang_CPL2021}, applied magnetic field\cite{Soh_PRB2019,YXu_PRB} and the chemical substitution\cite{Sanjeewa_PRB}. However, little is known about the doping induced effects on the electronic structure, magnetic structure and topological properties of this topological material.

Here we report the discovery of a giant and reversible electronic structure evolution in EuCd$_2$As$_2$. By high-resolution laser-based angle-resolved photoemission spectroscopy (ARPES) measurements, we directly observed the Fermi surface, and the band structure of EuCd$_2$As$_2$ experienced a dramatic change with doping that is induced by the surface absorption/desorption. We also observed a prominent band splitting at high doping corresponding to a doping-induced magnetic transition at low temperature (below $\sim$15\,K) from an antiferromagnetic state to a ferromagnetic state.

%Figure1
Initially, we found that the Fermi surface of EuCd$_2$As$_2$ exhibits a dramatic change with time, as shown in Fig. 1. The corresponding band structure evolution with time is shown in Fig. 2 (More complete Fermi surface and band structure evolutions with time are shown in Fig. S1 in Supplementary Materials\cite{SM}). The drastic Fermi surface change with time shows up in terms of the shape, the size and the number of Fermi surface sheets (Fig. 1c). The schematic Fermi surface for several typical dwell times, based on the detailed Fermi surface mapping analysis (Fig. S2 in Supplementary Materials\cite{SM}) and the band structure analysis, is shown in Fig. 1d. At the beginning, when EuCd$_2$As$_2$ is cleaved and measured at 250\,K, the measured Fermi surface is a point at $\bar\mathrm\Gamma$ (first panel in Fig. 1c). After the sample was quickly cooled down to 3\,K and immediately measured, the Fermi surface remains a tiny circle (second panel in Fig. 1c and first panel in Fig. 1d). The tiny hole pocket increases in its size with the dwell time up to 3 hours (third panel in Fig. 1c and second panel in Fig. 1d). At the dwell time of 6 hours, the observed Fermi surface consists of two pockets: a smaller hexagon-shaped sheet and another larger petal-like sheet (fourth panel in Fig. 1c). These two hole-like Fermi surface sheets get larger with time up to the dwell time of 12 hours (sixth panel in Fig. 1c and third panel in Fig. 1d). Further Fermi surface splitting occurs with time and at the time of 24 hours, four Fermi surface sheets can be clearly observed (ninth panel in Fig. 1c and fourth panel in Fig. 1d). These four Fermi surface sheets increase in their size with time and at the dwell time of 81 hours, the outer two Fermi surface sheets get so large that they are approaching the measurement window (second last panel in Fig. 1c and last panel in Fig. 1d). After that, when the sample is warmed up to 250\,K, the measured Fermi surface (last panel in Fig. 1c) recovers to a single point at $\bar\mathrm\Gamma$, similar to the Fermi surface measured at 250\,K before the cycle (first panel in Fig. 1c). When the sample is quickly cooled down again to 3\,K, similar dramatic evolution of the Fermi surface with time can be observed. The Fermi surface can always be recovered to a single point at $\bar\mathrm\Gamma$ whenever the sample is warmed up to 250\,K. The observed Fermi surface evolution with time at 2$\sim$3\,K was repeated several times on different samples. We find that the trend of the electronic structure evolution with time is the same but the speed of the evolution process varies between different samples and between different cycles even for the same sample.

%Figure2
Figure 2 shows the corresponding band structure evolution of EuCd$_2$As$_2$ with time measured at 3\,K. The band structure also exhibits a significant change with time in terms of the band position and the number of the observed bands. When the sample was cleaved and measured at 250\,K, two main bands are observed in the measurement energy window (marked as $\alpha$ and $\beta$ bands in the first panel of Fig. 2a). After the sample was quickly cooled down to 3\,K, the overall band structure shifts upward by $\sim$70\,meV, accompanied by the splitting of the $\alpha$ band (second panel in Fig. 2a). Till the dwell time of 6 hours, the two split bands are fully separated (marked as B1 and B2 bands in the fourth panel of Fig. 2a). When the dwell time reaches 15 hours, the original B1 and B2 bands start to further split and at the time of 24 hours, four well separated bands can be clearly observed (marked as B3, B4, B5 and B6 bands in the ninth panel of Fig. 2a). These four hole-like bands increase in their Fermi momenta with the further increase of the dwell time and after 39 hours, the outer band (B3) gets out of the detection window. On the other hand, the $\beta$ band experiences a more complicated evolution with the dwell time. Before the dwell time of 9 hours, this single $\beta$ band lowers its energy position with the increasing time by $\sim$100\,meV (second to fifth panels in Fig. 2a). After 9 hours, the energy position of this $\beta$ band shifts continuously upwards by $\sim$260\,meV up to 81 hours. The $\beta$ band also starts to split after the dwell time of 18 hours and eventually two well-split bands can be observed at 81 hours (marked as B7 and B8 bands in the last second panel of Fig. 2a). When the sample was warmed up to 250\,K again, the two original $\alpha$ and $\beta$ bands can be recovered as seen in the last panel in Fig. 2a.

In order to keep track on the band structure evolution with the dwell time, in Fig. 2d and 2e, we plot the change of momentum distribution curves (MDCs) at the Fermi level and at the binding energy of 100\,meV, respectively. From these figures, the initial splitting of the original $\alpha$ band into B1 and B2 bands, further splitting of the B1 and B2 bands into four B3-B6 bands, and the continuous Fermi momentum increase of these bands with the increasing time can be clearly visualized. The splitting of the original $\beta$ band into B7 and B8 bands and the corresponding momentum increase with the increasing time can also be clearly observed in Fig. 2e. The energy position change of the $\beta$ band can be clearly seen from the photoemission spectra (energy distribution curves, EDCs) at the $\bar\mathrm\Gamma$ point in Fig. 2f. The top of the $\beta$ band exhibits a non-monotonic variation with time: it first shifts to higher binding energy during the dwell time of 0$\sim$9 hours and then moves to lower binding energy in the time range of 9$\sim$81 hours. We notice that, during the time of 0$\sim$9 hours, the $\alpha$ and $\beta$ bands shift with time along the opposite energy direction. This clearly indicates that the band shift with the dwell time is not a rigid band shift.

We have also carried out electronic structure evolution experiments on EuCd$_2$As$_2$ at 25\,K (Fig. S3 in Supplementary Materials\cite{SM}) and 150\,K (Fig. S4 in Supplementary Materials\cite{SM}). At the temperature of 25\,K which is higher than the antiferromagnetic transition temperature of EuCd$_2$As$_2$ at 9.3\,K, obvious electronic structure evolution with time is also observed. It involves both the Fermi surface topology change and the band structure position shift (Fig. S3), similar to those we observed at 3\,K (Fig. 1 and Fig. 2). Even at 150\,K, the electronic structure evolution can still be observed (Fig. S4). In this case, it mainly involves the energy position shift of the two $\alpha$ and $\beta$ bands.

%Figure3
We have measured the electronic structure of the fully-modified EuCd$_2$As$_2$ at different temperatures. This is realized by keeping the EuCd$_2$As$_2$ sample for a sufficiently long time at a low temperature and then slowly changing to other temperatures; at each temperature, the sample stayed for 3 hours when the ARPES measurement was carried out. Fig. 3 shows the measured Fermi surface and band structure at several typical temperatures in the low temperature regime between 2$\sim$30\,K; another independent measurement in a larger temperature range between 3$\sim$250\,K is shown in Fig. S5 in Supplementary Materials\cite{SM}. For the fully-modified EuCd$_2$As$_2$, its electronic structure exhibits a dramatic change with temperature, particularly in the low temperature range between 2$\sim$30\,K.

For the modified sample at 2\,K, the band splitting is fully realized with four bands (B3, B4, B5 and B6) clearly observed near the Fermi level (first panel in Fig. 3b). This band splitting persists up to 9.3\,K, the antiferromagnetic transition temperature of EuCd$_2$As$_2$ (second panel in Fig. 3b). Above 9.3\,K, the intensity of the B3, B4 and B5 bands quickly gets weaker and  becomes nearly invisible above 20\,K (25\,K and 30\,K panels in Fig. 3b). Correspondingly, the Fermi surface topology changes from four Fermi surface sheets at 2\,K and 9.3\,K to a single main sheet at 25\,K and 30\,K as seen from Fig. 3a.

From our detailed temperature-dependent measurements in a wide temperature range of 3$\sim$250\,K (Fig. S5 and Fig. S6 in Supplementary Materials\cite{SM}), we find that there are three distinct temperature regions where the electronic structures of the modified EuCd$_2$As$_2$ behave quite differently. Fig. 3c shows the stacked second-derivative MDCs at the binding energy of 25\,meV from the band structures measured at different temperatures (Fig. S5b in Supplementary Materials\cite{SM}). The stacked second-derivative MDCs at other binding energies (Fig. S6 in Supplementary Materials\cite{SM}) give similar results. In the temperature region I of 3$\sim$30\,K, as described before, the electronic structure exhibits a dramatic change with temperature in terms of the band position and the number of the observed bands. In the second temperature region II of 30$\sim$135\,K, the observed electronic structure shows little change with temperature, keeping similar band position and the same number of bands. When moving to the third temperature region III of 135$\sim$250\,K, the measured electronic structure shows a significant change with temperature again. The position of the B1 band and the $\beta$ band keeps shrinking with the increasing temperature.

In order to keep track on the band position change with temperature, we show in Fig. 3e the EDCs at the $\bar\mathrm\Gamma$ point measured on the modified EuCd$_2$As$_2$ at different temperatures obtained from the measurements in Fig. S5. Two EDC peaks are observed, originating from the $\gamma$ band and $\delta$ band, as shown in Fig. S7 in Supplementary Materials\cite{SM}. In the temperature range of 2$\sim$30\,K, the position of these two bands moves toward the Fermi level with increasing temperature. Their positions exhibit little change with temperature in the range of 30$\sim$135\,K. Upon further increase of the temperature, the position of these two bands shifts to higher binding energies. The temperature evolution of the EDCs at $\bar\mathrm\Gamma$ is consistent with the MDC evolution in Fig. 3c in terms of three distinct temperature regions. We note that the overall band shift with temperature is not monotonic. Furthermore, it is not a rigid band shift because the magnitude of the band position shift between the $\gamma$ band and the $\delta$ band is different and the number of the observed bands changes significantly at different temperatures.

%%%Discussion

%0.XRD,STM,LEED
The observation of such a dramatic electronic structure evolution with time in EuCd$_2$As$_2$ is rare. The first question to ask is whether this electronic structure change is associated with the sample surface or the entire bulk. To this end, we measured the time-dependence of the resistivity and heat capacity of EuCd$_2$As$_2$ at different temperatures and these bulk properties do not exhibit obvious change with time even at a low temperature of 5\,K\cite{YWang_CPL2021}. This is in a strong contrast to the dramatic electronic structure evolution we have observed. It indicates that the observed electronic structure change is mainly associated with the sample surface since ARPES is a surface sensitive technique. The second question is whether there is a structure change with time in EuCd$_2$As$_2$. The crystal structure of EuCd$_2$As$_2$ exhibits little change with time from the time-dependent X-ray diffraction (XRD) measurement at a low temperature\cite{YWang_CPL2021}. By performing time-dependent scanning tunneling microscope (STM) (Fig. S8 in Supplementary Materials\cite{SM}) and low energy electron diffraction (LEED) (Fig. S9 in Supplementary Materials\cite{SM}) measurements on EuCd$_2$As$_2$, we find that there is no surface structure change with time at low temperature. These results demonstrate that the electronic structure evolution observed  in EuCd$_2$As$_2$ is not due to bulk or surface structure change with time.

%Figure4  
The measured Fermi surface (Fig. 1c) and band structure (Fig. 2) of EuCd$_2$As$_2$ indicate that, during the evolution process, there is a continuous increase of the charge carriers (holes) in the measured region. When EuCd$_2$As$_2$ is fully modified, the hole doping level can reach up to $\sim$0.18 holes per unit cell (Fig. S10 in Supplementary Materials\cite{SM}). A natural question to ask on the origin of the dramatic electronic structure change is where the charge carriers come from. There are two main possibilities. The first is related to the intrinsic relaxation process in EuCd$_2$As$_2$, such as the long-time magnetic relaxation with time\cite{YWang_CPL2021} or some other unidentified relaxation processes that causes such a dramatic electronic structure evolution with time. The second possibility is related to an extrinsic surface absorption/desorption process that causes a dramatic change of charge carriers at the surface. This process can be further divided into a natural absorption or a laser-assisted absorption where the laser may cause or enhance the surface absorption\cite{RJiang_PRB2012}. To pin down on the exact origin, we have taken the measurements many times under different experimental conditions. We find that, although the trend of the electronic structure evolution with time is the same, the speed of the evolution varies between different samples and between different cycles even for the same sample. In particular, we found that the electronic structure evolution speed is closely related to the vacuum level in the ARPES measurement chamber. Fig. 4 shows the electronic structure change of EuCd$_2$As$_2$ with time measured under different experimental conditions on the same sample. In this case, when the vacuum is good and the laser is off in between the measurements, the sample experiences little electronic structure change with time even after $\sim$27 hours, as shown in Fig. 4a and 4f. This is in a strong contrast to the results in Fig. 1 where a giant electronic structure change already occurs after the similar time duration. This difference can basically rule out the first possibility that the dramatic electronic structure evolution with time observed in Fig. 1 is associated with the intrinsic relaxation process in EuCd$_2$As$_2$, making the surface absorption a more likely origin. When we maintained all the measurement conditions the same, but only kept the laser always on the sample, we found that the electronic structure started to show an obvious change with time, as shown in Fig. 4b and 4g. This indicates that the laser played a role in the electronic structure evolution process and it can be related to the laser-assisted absorption\cite{RJiang_PRB2012}. To confirm the surface absorption, we warmed up the sample to 250\,K to let its electronic structure recovered to its initial state. In the meantime, the vacuum level in the chamber got worse during this warming-up process. After the sample was quickly cooled down to 5\,K, even when the laser was turned off in between the measurements like the situation in Fig. 4a, the electronic structure experiences an obvious change with time, as shown in Fig. 4d and 4i. These measurements indicate that a natural absorption process can also give rise to a dramatic electronic structure variation when the vacuum level is relatively lower. When the natural absorption process already produces a substantial change in the electronic structure, the laser-assisted absorption may further slightly enhance the process, but its effect may become less dominant, as shown in Fig. 4e and 4j. The results in Fig. 4 clearly indicate that the dramatic electronic structure evolution we observed in Fig. 1 can be mainly attributed to the surface absorption/desorption process. Since hydrogen is the major ingredient of the residual gas in our ultra-high vacuum chamber, we suspect the charge carrier change on surface is mainly due to hydrogen absorption/desorption although further work needs to be done to pin down on the exact origin.

%Figure5
In order to understand the dramatic electronic structure evolution with hole doping (Fig. 1 and Fig. 2) and the temperature evolution of the modified EuCd$_2$As$_2$ (Fig. 3), we carried out detailed electronic structure calculations of EuCd$_2$As$_2$ bulk (Fig. 5, Fig. S11 and Fig. S12 in Supplementary Materials\cite{SM}),  EuCd$_2$As$_2$ slab (Fig. S13 in Supplementary Materials\cite{SM}) and EuCd$_2$As$_2$ monolayer (Fig. S14 in Supplementary Materials\cite{SM}) in different magnetic states. For the undoped EuCd$_2$As$_2$ in the paramagnetic state at 250\,K, ARPES should measure the bulk electronic structure and the surface states if they exist. When only the top layers are doped due to the surface absorption, it may be more appropriate to understand the electronic structure from the band structure calculations of the monolayers. In Fig. S15 in Supplementary Materials\cite{SM}, we showed the comparison between the measured band structure of EuCd$_2$As$_2$ at 250\,K and the calculated results of EuCd$_2$As$_2$ monolayer, bulk and slab in the paramagnetic state. All the three calculated results are similar in that they consist of three main branches of bands labeled as B1, B2 and $\beta$ within the energy range of $\sim$1\,eV below the Fermi level (Fig. S15b, S15c and S15d). The measured band structure is consistent with the band structure calculations in that three bands are observed labeled as $\alpha$ and $\beta$ with $\alpha$ consisting of two bands (B1 and B2 in Fig. S15b in Supplementary Materials\cite{SM}). However, we find that neither of these calculations can be fully consistent with the measured results in the quantitative manner. This deviation may be due to some correlation effects that are not properly considered in the calculations and further theoretical efforts are needed to address this issue.

Fig. 5 highlights the calculated electronic structures of EuCd$_2$As$_2$ and their comparison with the doping evolution of the  measured electronic structures. In the paramagnetic state, there are mainly three bands (B1, B2 and $\beta$) near the Fermi level (Fig. 5n and 5o); the corresponding Fermi surface is only a point (Fig. 5m). The measured band structures (Fig. 5b and 5c) and the Fermi surface (Fig. 5a) of EuCd$_2$As$_2$ at 250\,K in the paramagnetic state are qualitatively consistent with the calculated results although the bands B1 and B2  are nearly degenerate ($\alpha$ band in Fig. 5b and 5c) in this case. In the initial stage of doping evolution, the measured bands (Fig. 5e and 5f) exhibit a slight upward shift and an apparent B1 and B2 splitting; the overall band structure and Fermi surface (Fig. 5d) are still consistent with the calculated results in the paramagnetic state (Fig. 5m, 5n and 5o). In the early stage of evolution at 2$\sim$3\,K, the measured band structures (Fig. 5h and 5i) show an obvious upward shift and the B1 and B2 bands become well separated. The measured Fermi surface (Fig. 5g) consists of two pockets. These results are in a good agreement with the calculations (Fig. 5p, 5q and 5r) when the chemical potential is shifted by $\sim$0.3\,eV. In the later stage of doping evolution at 2$\sim$3\,K, 6 bands appear near the Fermi level (B3-B8 in Fig. 5k and 5l) and 4 Fermi surface sheets are observed (Fig. 5j). Such electronic structures can not be understood if EuCd$_2$As$_2$ is in the paramagnetic state or in the antiferromagnetic state (Fig. S11 in Supplementary Materials\cite{SM}). On the other hand, these electronic structures are rather consistent with the calculated results with EuCd$_2$As$_2$ being in the ferromagnetic state and an overall upward band shift of $\sim$0.4\,eV (Fig. 5v, 5w and 5x). The analysis of this electronic structure evolution indicates that, during the doping evolution process, in addition to the significant band shift, there is also a change of the magnetic state at low temperature from the initial antiferromagnetic state at zero or low doping to ferromagnetic state at high doping. This is consistent with the recent observation of band filling controlled magnetism in EuCd$_2$As$_2$: the antiferromagnetic state is observed at low doping while the ferromagnetic state is observed at high doping\cite{Jo_PRB}.

%Temperature evolution
The particular electronic structure evolution with temperature observed in the fully modified EuCd$_2$As$_2$, as shown in Fig. 3, can be understood by combining the doping change and the magnetic state change. In the temperature region I of 3$\sim$30\,K, the electronic structure exhibits a dramatic change with temperature in both the band position and the number of the bands. Although the hole doping may be gradually decreased due to surface desorption with increasing temperature, such a doping change alone is hard to understand the drastic electronic structure change in this temperature region. The magnetic state of EuCd$_2$As$_2$ at low temperatures most likely plays an important role in dictating its electronic structure. Coincidentally, this is the temperature range where below $\sim$30\,K the magnetic susceptibility shoots up and the sample enters into the magnetic states, as shown in Fig. 3d. In the temperature region II of 30$\sim$135\,K, the band structure shows little change with temperature. Above 30\,K, EuCd$_2$As$_2$ enters into a paramagnetic state and its electronic structure is mainly controlled by the doping level. The increasing temperature tends to decrease the absorption but, in the meantime, it causes the vacuum level in the ARPES chamber to become worse which tends to increase the absorption. Therefore, the doping shows little change with temperature in this temperature region because of these two competing absorption/desorption processes involved. In the temperature region III above $\sim$135\,K, the surface desorption process due to the temperature increase becomes dominant. The doping level keeps decreasing with increasing temperature until nearly negligible surface absorption is left and the EuCd$_2$As$_2$ sample recovers to its initial non-doped state at $\sim$250\,K. The reversible electronic structure evolution can thus be understood by this surface absorption/desorption process. 

%Compare HDing %Doping induced magnetic change\cite{Jo_PRB}
Our present results provide a new perspective in understanding the previous ARPES results on EuCd$_2$As$_2$\cite{JZMa_SA,Soh_PRB2019,JZMa_AM,Jo_PRB}. Our results clearly indicate that, when EuCd$_2$As$_2$ is undoped or only slightly doped, the electronic structure in the paramagnetic state (250\,K) (Fig. 4c and 4h) is similar to that observed at low temperature in the antiferromagnetic state ($<$9.3\,K) (Fig. 4a and 4f). It is consistent with the observations reported in Ref.\cite{Soh_PRB2019} but quite different from the results in Ref.\cite{JZMa_SA}. It was reported in Ref.\cite{JZMa_SA} that the degeneracy of the Bloch bands is already lifted in the paramagnetic state of EuCd$_2$As$_2$ which is attributed to the presence of the quasi-static and quasi-long-range ferromagnetic fluctuations. Our high resolution laser ARPES measurements demonstrate unambiguously that such a lifting of the band degeneracy does not occur in the entire measured temperature range of 2$\sim$250\,K. On the other hand, our present work asks for special caution to be exercised about surface absorption in ARPES measurements of EuCd$_2$As$_2$ which might not be noticed before. The band structures observed in Ref.\cite{JZMa_SA} are similar to those we observed in EuCd$_2$As$_2$ with sufficient hole doping due to surface absorption. These indicate that the band splitting observed in Ref.\cite{JZMa_SA} is more likely due to hole doping from surface absorption rather than the ferromagnetic fluctuation as they proposed. 

%Summary
In summary, we have discovered a giant and reversible electronic structure evolution phenomenon in EuCd$_2$As$_2$. A large amount of hole doping is introduced into the sample surface during the cooling process because of surface absorption that dramatically changes the electronic structures. The doping also causes a magnetic state transition at low temperature (below $\sim$15\,K) from an antiferromagnetic state at zero or low dopings to a ferromagnetic state at high doping which give rise to band splitting in EuCd$_2$As$_2$. The discovery of the dramatic electronic structure evolution process in EuCd$_2$As$_2$ due to surface absorption asks to reexamine the previous results\cite{JZMa_SA,JZMa_AM,Jo_PRB}. Our results have established a detailed electronic phase diagram of EuCd$_2$As$_2$ where the electronic structure and the magnetic structure change systematically and dramatically with the hole doping. They further suggest that the transport, magnetic and topological properties of the surface or bulk EuCd$_2$As$_2$ can be greatly modified. There are many feasible ways to change the doping like aliovalent substitution, formation of vacancies or interstitials\cite{Jo_PRB}, or gating. The present work will motivate people to explore for new phenomena and properties by doping the bulk EuCd$_2$As$_2$ or EuCd$_2$As$_2$ thin films.\\

\noindent {\bf Methods}\\
\noindent{\bf Sample} Single crystals of EuCd$_2$As$_2$ were grown by Sn flux method\cite{HPWang_PRB}. High purity elements of Eu, Cd, As, and Sn were put in an alumina crucible at a molar ratio of 1:2:2:10 and sealed in a quartz tube under high vacuum. The tube was heated to 1173\,K, kept for 20 hours, and then slowly cooled to 773\,K at a rate of 2\,K/hour. After that the single crystals were separated from the Sn liquid in a centrifuge. \\

\noindent{\bf ARPES Measurements} High-resolution ARPES measurements were performed at our laboratory-based systems equipped with the 6.994\,eV vacuum ultraviolet laser and the time-of-flight electron energy analyzer (ARToF 10k by Scienta Omicron) or a hemisphere analyzer (DA30L by Scienta Omicron)\cite{GDLiu_RSI2008,XJZ_Review}. The ARToF-ARPES system can cover two-dimensional momentum space ($k_x$, $k_y$) simultaneously. It is also equipped with an ultra-low temperature cryostat which can cool the sample down to a low temperature of 1.6\,K. The overall energy resolution is $\sim$1\,meV and the angular resolution is $\sim$0.1 degree. The Fermi level is referenced by measuring polycrystalline gold which is in good electrical contact with the sample. The samples were cleaved {\it in situ} and measured at different temperatures in ultrahigh vacuum with a base pressure better than 5.0$\times$10$^{-11}$\,mbar. \\

\noindent{\bf STM Measurements} Scanning tunneling microscope measurements were performed at the home-made ultra-high vacuum (UHV) low temperature STM system. The samples were cleaved at 4.2\,K. During the measurements, the current and the V-bias were set at $\sim$100\,pA and $\sim$100\,mV, respectively. The scanning range is about 30\,nm $\times$ 30\,nm.\\

\noindent{\bf Band Structure Calculations} Band structure calculations were performed on EuCd$_2$As$_2$ bulk, EuCd$_2$As$_2$ slab and EuCd$_2$As$_2$ monolayer. For the bulk calculations, we calculated the electronic structure of EuCd$_2$As$_2$ in 5 different magnetic states, i.e., the paramagnetic state, the ferromagnetic state with the in-plane and out-of-plane magnetizations, and the inter-layer anti-ferromagnetic state with the in-plane and out-of-plane magnetizations. For the antiferromagnetic state, we made a 1$\times$1$\times$2 supercell to get consistent with the c axis magnetic order. We calculated the electronic structure with the Vienna $ab$ $initio$ Simulation package (VASP)\cite{GKresse_PRB1996,John_PRL1996}. The projector-augmented-wave (PAW) method\cite{PEBlochl_PRB1994,GKresse_PRB1999} with the Perdew-Burke-Ernzerhof (PBE) exchange-correlation\cite{John_PRL1996} functional was used. In the paramagnetic state, we did not take the $f$ electrons of Eu in our pseudo-potential. For the ferromagnetic and the antiferromagnetic state, we took the $f$ electrons into consideration and added the on-site $U$ as 5\,eV. We set a 13$\times$13$\times$7 $k$-mesh for the primitive unit cell (paramagnetic state and ferromagnetic state) and a 13$\times$13$\times$4 $k$-mesh for the supercell (antiferromagnetic state). In all cases, we set the plane wave energy cutoff as 500\,eV. We also calculated the constant energy contours at different binding energies. For these calculations, we constructed tight-binding Hamiltonians using the WANNIER90 package\cite{GPizzi_JPCM2020}, and calculated the constant energy contours by WANNIERTOOLS package\cite{QSWu_CPC2018}. For the slab calculations, we constructed a 1$\times$1$\times$8 supercell and add a vacuum layer larger than 15\,$\AA$. For the monolayer calculations, the Vienna $ab$ $initio$ simulation package (VASP) based on the density functional theory (DFT) was used. The vacuum layer of 20\,$\AA$ along the (001) direction was used to avoid the interaction between nearest monolayers. The Perdew-Burke-Ernzerhof (PBE) exchange-correlation function with the generalized gradient approximation (GGA) $+$ $U$ method was used to treat the localized 4$f$ orbitals of Eu atom\cite{John_PRL1996,HJKulik_PRL2006}, and the $U$ parameter was selected to be 5\,eV. The cutoff energy was set at 500\,eV and the Brillouin zone (BZ) integral was sampled by 9$\times$9$\times$1 $k$-mesh. The spin-orbital coupling (SOC) was taken into account in all calculations.\\

\vspace{3mm}

\noindent {\bf Acknowledgement}\\
We thank Changjiang Yi for the EuCd$_2$As$_2$ crystal growth and  Xianxin Wu for the theoretical discussions. We thank financial support from the National Natural Science Foundation of China (Nos. 11888101, 11974404, 11674369, 11925408, U2032204 and 11227903), the National Key Research and Development Program of China (Nos. 2021YFA1401800, 2017YFA0302900, 2017YFA0302901, 2017YFA0302903, 2018YFA0305600 and 2018YFA0305700), the Strategic Priority Research Program (B) of the Chinese Academy of Sciences (No. XDB33000000 and XDB28000000), the Youth Innovation Promotion Association of CAS (No. 2017013), the Beijing Natural Science Foundation (No. Z180008), Beijing Municipal Science and Technology Commission (No. Z191100007219013, Z181100004218007 and Z191100007219011), the Research Program of Beijing Academy of Quantum Information Sciences (No. Y18G06), the K. C. Wong Education Foundation (No. GJTD-2018-01) and the Synergetic Extreme Condition User Facility (SECUF).

\vspace{3mm}

\noindent {\bf Author Contributions}\\
Y.W., C.L., T.M.M., S.Z. and Y.L. contributed equally to this work. X.J.Z., Y.W. and C.L. proposed and designed the research. Y.L. and M.Y. contributed to EuCd$_2$As$_2$ crystal growth. Y.W., C.L., T.M.M., C.H.Y., Y.Q.C., C.Y.S., H.L.L., H.C., L.Z., F.F.Z., F.Y., Z.M.W., S.J.Z., Q.J.P., Z.Y.X. and X.J.Z. contributed to the development and maintenance of Laser ARPES systems. S.Z., L.Q.Z., H.M.W. and T.X. contributed to the band structure calculations and theoretical discussions. H.B.D., Y.K.S., C.J.Z., H.Q.M. and S.H.P. contributed to STM measurements. Y.W., C.L. and T.M.M. carried out the ARPES experiment with the assistance from C.H.Y., Y.Q.C., C.Y.S., H.L.L. and H.C.. Y.W., C.L., T.M.M. and X.J.Z. analyzed the data. Y.W., C.L. and X.J.Z. wrote the paper with H.M.W. and T.X.. All authors participated in discussion and comment on the paper.

\newpage

\begin{figure*}[tbp]
\begin{center}
\includegraphics[width=0.8\columnwidth,angle=0]{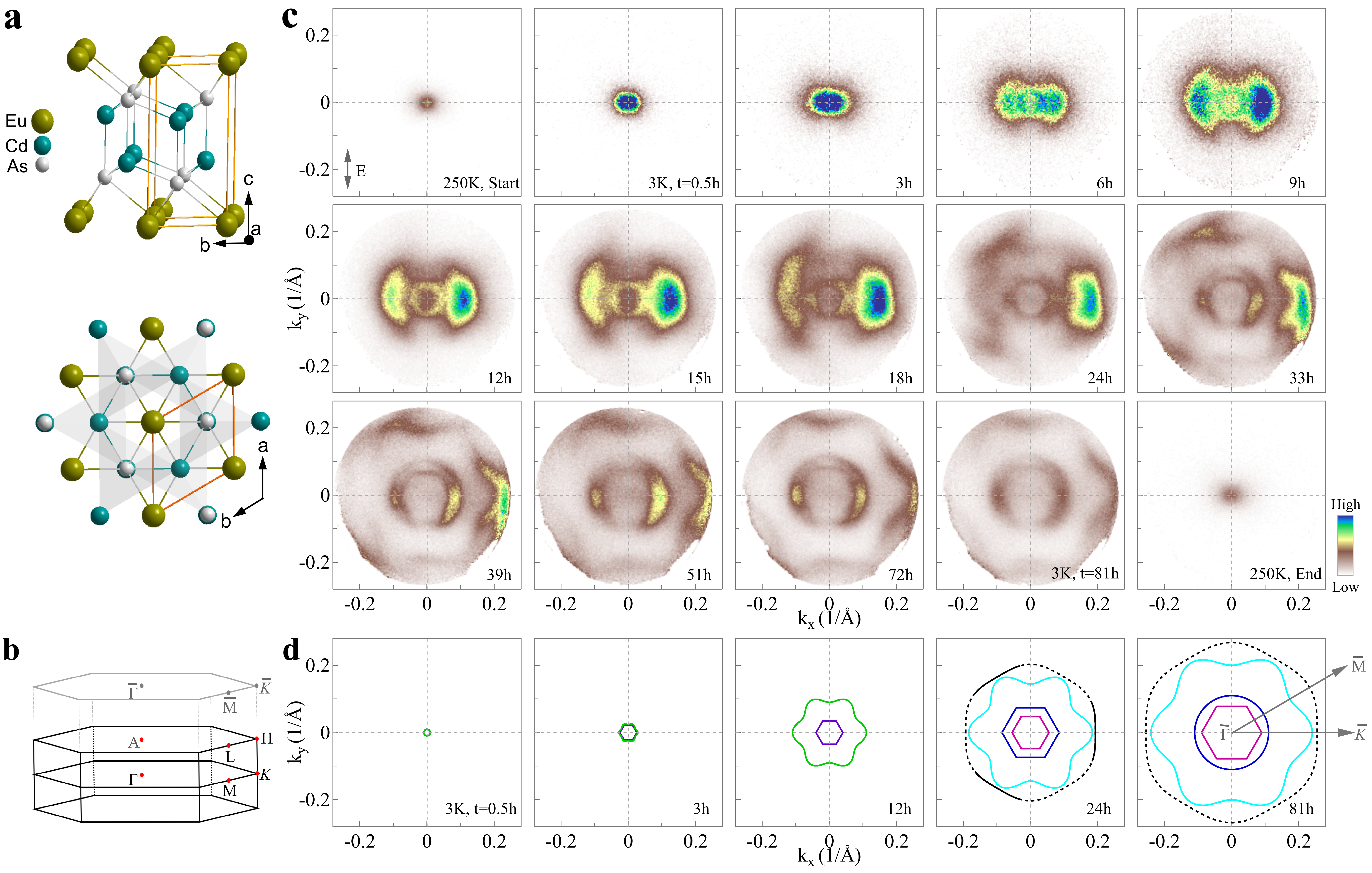}
\end{center}
\caption{\textbf{Giant and reversible Fermi surface evolution with time of EuCd$_2$As$_2$ at 3\,K.} (a) The crystal structure of EuCd$_2$As$_2$ with a space group $\mathrm{P\bar{3} m1}$. The $\mathrm{Cd_2As_2}$ bilayers are sandwiched in between two adjacent Eu layers. The upper panel represents a 3-dimensional crystal structure where the unit cell is indicated by the orange lines. The lower panel shows the projection of the 3-dimensional crystal structure along the c axis. The orange rhombus represents the 2-dimensional unit cell. (b) Three-dimensional Brillouin zone (BZ) of EuCd$_2$As$_2$ (black solid lines) and the corresponding two-dimensional BZ projected on the (001) plane (gray solid lines). The high symmetry points are indicated. (c) Fermi surface of EuCd$_2$As$_2$ measured on the (001) surface at 3\,K at different dwell times. The sample was first cleaved at 250\,K with its Fermi surface measured (first panel in (c)). Then the sample was quickly cooled down to 3\,K and the Fermi surface was measured at 3\,K at different times. The laser is always on the sample surface during the entire measurements. The Fermi surface was continuously measured up to 81 hours and a dramatic change of the Fermi surface with time is observed. Then the sample was warmed up to 250\,K and the Fermi surface was measured at 250\,K again (last panel in (c)). Detailed Fermi surface evolution with time is shown in Fig. S1a in Supplementary Materials\cite{SM}. The polarization of the laser light is marked by a double arrow in the first panel. (d) Schematic Fermi surface of several representative panels in (c) during the evolution process. When the dwell time is longer than 24 hours (last two panels in (d)), some parts of the outer Fermi surface sheets are out of the detection window and are marked by dashed lines.}
\end{figure*}

\begin{figure*}[tbp]
\begin{center}
\includegraphics[width=1\columnwidth,angle=0]{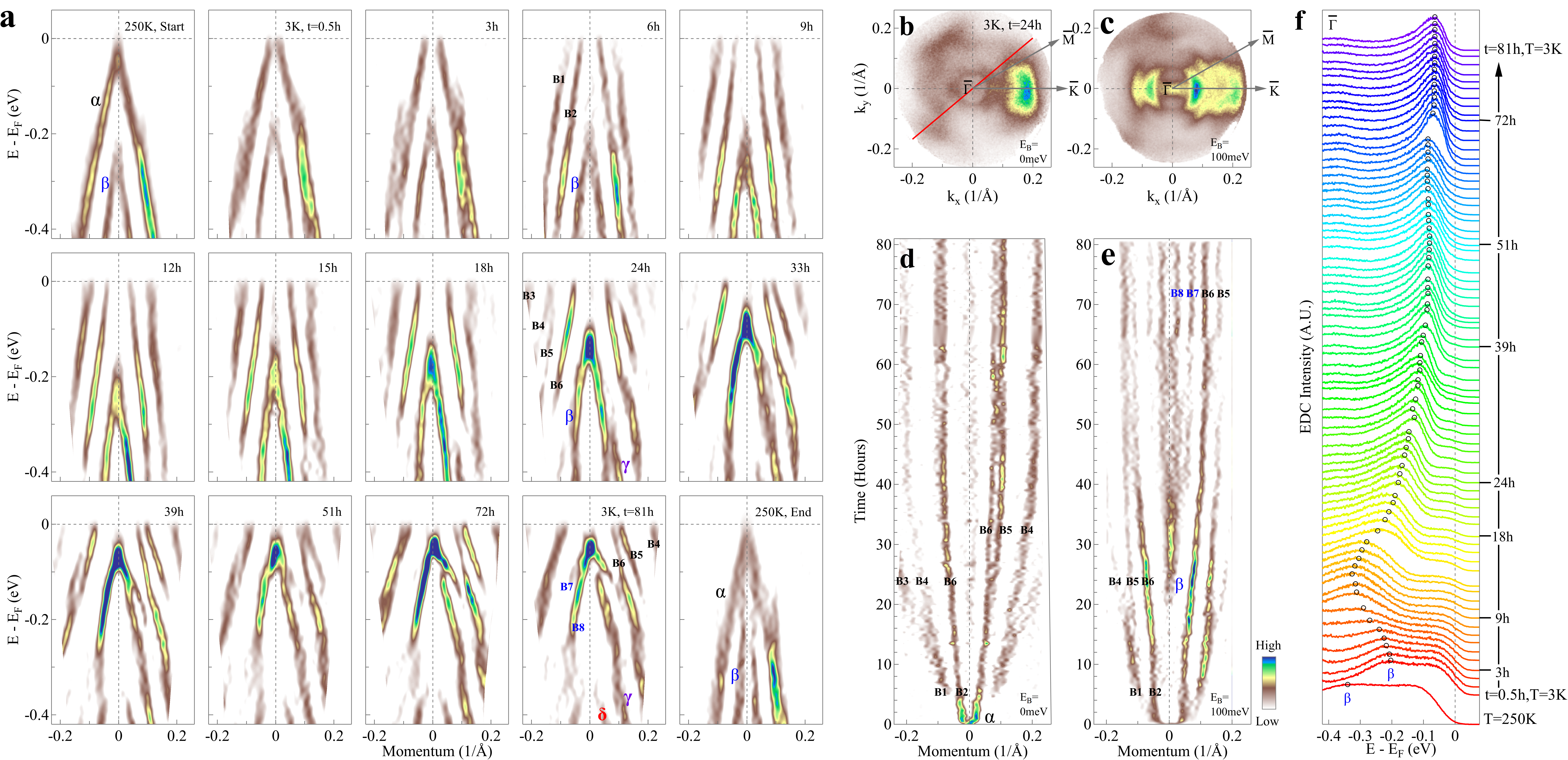}
\end{center}
\caption{\textbf{Giant and reversible band structure evolution with time of EuCd$_2$As$_2$ at 3\,K.} (a) Band structure of EuCd$_2$As$_2$ measured at 3\,K at different dwell times. The bands are obtained from the same measurements in Fig. 1c. The location of the momentum cut is marked in (b) by a red line. These are the second derivative images with respect to the momentum. The observed bands are labeled by $\alpha$, $\beta$ , $\gamma$, $\delta$ and B1-B8. A significant change of the band structure is observed at different measurement times. It was found that the band structure at 250\,K after cooling at 3\,K (last panel in (a)) is similar to the one before cooling (first panel in (a)). (b-c) Fermi surface (b) and constant energy contour at the binding energy (E$_B$) of 0.1\,eV (c) after the dwell time of 24 hours at 3\,K. Detailed band structure evolution with time is shown in Fig. S1b in Supplementary Materials\cite{SM}. (d-e) Two-dimensional image showing the momentum distribution curves (MDCs) at different measurement times. The MDCs are obtained from (a) and Fig. S1 at the Fermi level (d) and the binding energy of E$_B$=0.1\,eV (e). The observed bands are labeled by $\alpha$, $\beta$ and B1-B8. (f) Photoemission spectra (energy distribution curves, EDCs) measured at 3\,K at the $\bar\mathrm\Gamma$ point after different dwell times. On each EDC, the energy position of the $\beta$ band (first panel in (a)) is marked by an empty circle.  
}
\end{figure*}

\begin{figure*}[tbp]
\begin{center}
\includegraphics[width=1\columnwidth,angle=0]{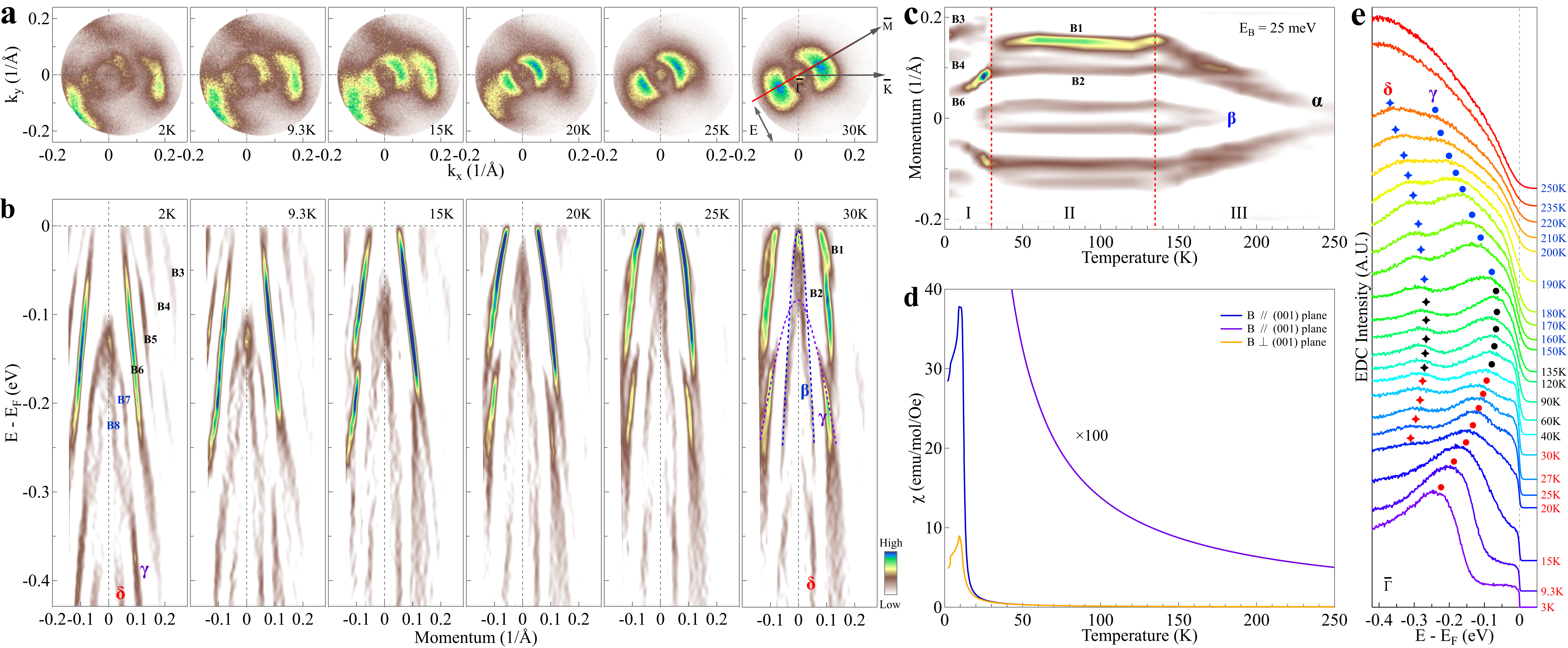}
\end{center}
\caption{\textbf{Temperature-dependent electronic structure of the modified EuCd$_2$As$_2$.} (a) Fermi surface of the modified EuCd$_2$As$_2$ measured at different temperatures. The sample was first kept at 30\,K for a long time to make its electronic structure nearly fully modified. Then the sample was cooled down to different temperatures of 25\,K, 20\,K, 15\,K, 9.3\,K and 2\,K. At each temperature, the sample stayed for 3 hours when the ARPES measurement was carried out. The polarization of the laser light is marked by a double arrow in the last panel. (b) Corresponding band structures measured at different temperatures. The location of the momentum cut is marked in the last panel of (a) by a red line. These are the second derivative images with respect to the momentum. Another independent and detailed measurement results over a wide temperature range (3$\sim$250\,K) are shown in Fig. S5 in Supplementary Materials\cite{SM}. The observed bands are labeled by B1-B8, $\beta$, $\gamma$ and $\delta$ in the first and last panels. (c) Two-dimensional image showing the second-derivative MDCs at different temperatures. The MDCs are obtained from Fig. S5 at the binding energy of E$_B$=0.025\,eV. The observed bands are labeled by $\alpha$, $\beta$ and B1-B6. The temperature evolution can be divided into three regions: 3$\sim$30\,K (I), 30$\sim$135\,K (II) and 135$\sim$250\,K (III). (d) Magnetic susceptibility of EuCd$_2$As$_2$ measured with the magnetic field parallel (blue line) and perpendicular (orange line) to the (001) plane. To highlight the magnetic susceptibility at high temperature, the original data (blue line) is multiplied 100 times and is shown in (d) by a purple line. (e) EDCs measured at different temperatures at the $\bar\mathrm\Gamma$ point. On each EDC, the energy position of the $\gamma$ band and $\delta$ band is marked by circle and asterisk , respectively.  
}
\end{figure*}

\begin{figure*}[tbp]
\begin{center}
\includegraphics[width=0.9\columnwidth,angle=0]{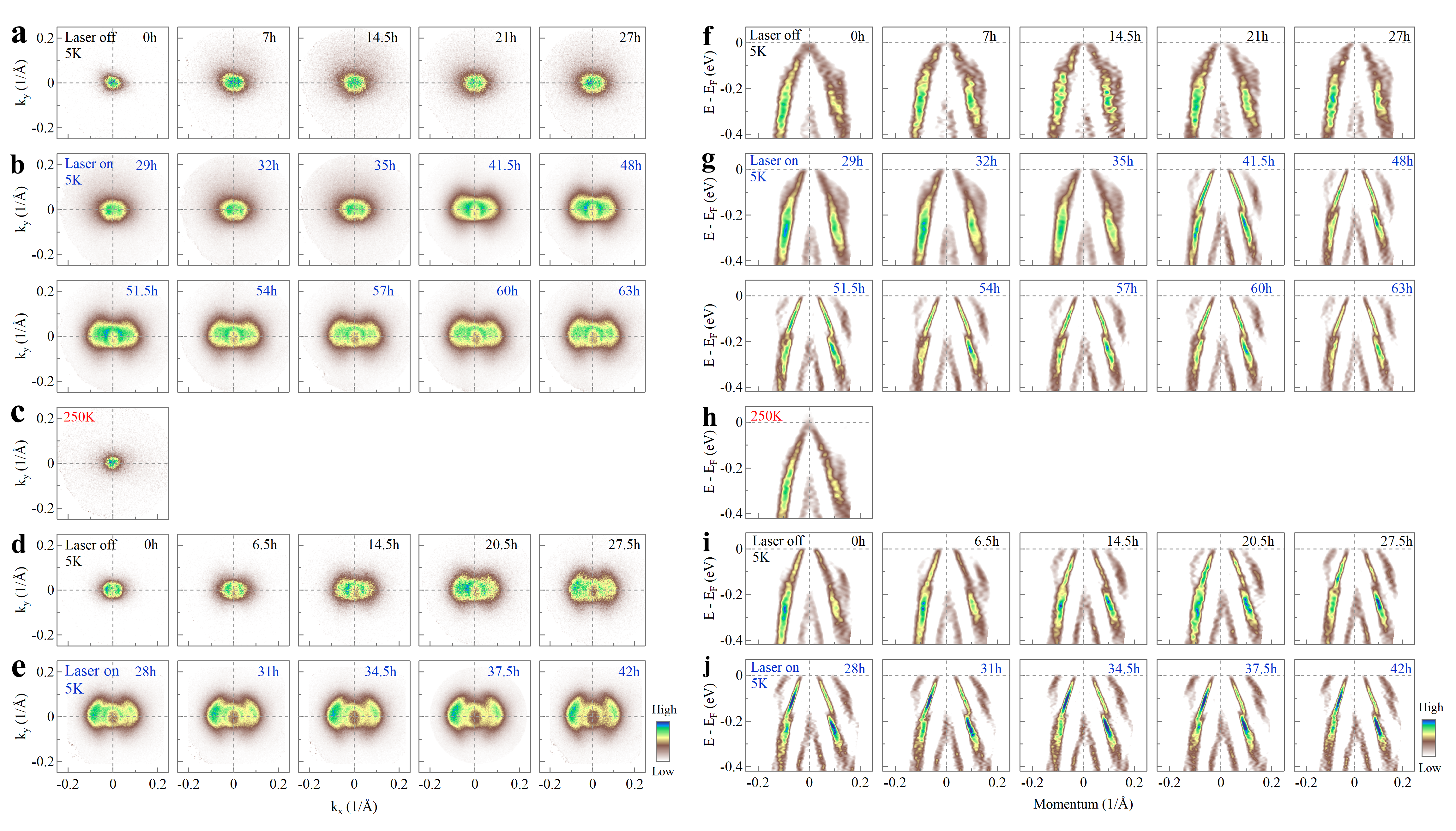}
\end{center}
\caption{\textbf{Electronic structure evolution of EuCd$_2$As$_2$ with time measured under different experimental conditions} (a) Fermi surface measured at 5\,K after different times marked at the upper-right corner of each panel. Each measurement took 30 minutes. The laser was turned off between the two consecutive measurements. The vacuum in the chamber is better than 5.0$\times$10$^{-11}$\,mbar. (b) Fermi surface measured at 5\,K after different times. The process continued after the measurements in (a) but, in this case, the laser was always on the sample during the measurements. Each measurement took 3 hours. The vacuum in the chamber remained better than 5.0$\times$10$^{-11}$\,mbar. (c) After the measurements in (b), the sample was warmed up to 250\,K and the vacuum in chamber went up to a peak pressure of 3.0$\times$10$^{-10}$\,mbar and then started to drop down. The Fermi surface was measured at 250\,K at a pressure of 1.0$\times$10$^{-10}$\,mbar. (d) The sample was quickly cooled down to 5\,K and the vacuum in the chamber became 6.0$\times$10$^{-11}$\,mbar and gradually recovered with time. The Fermi surface was measured immediately after the sample was cooled down to 5\,K and then measured after different times. Each measurement took 30 minutes. The laser was turned off between the two consecutive measurements. The vacuum in the chamber became better than 5.0$\times$10$^{-11}$\,mbar at the end of the measurements. (e) Fermi surface measured at 5\,K after different times. The process continued after the measurements in (d) but, in this case, the laser was always on the sample during the measurements. Each measurement took 3 hours. The vacuum in the chamber kept better than 5.0$\times$10$^{-11}$\,mbar. (f-j) The corresponding band structures measured in (a-e). The location of the momentum cut is marked by a red line in (a). These are the second derivative images with respect to momentum.
}

\end{figure*}

\begin{figure*}[tbp]
\begin{center}
\includegraphics[width=1\columnwidth,angle=0]{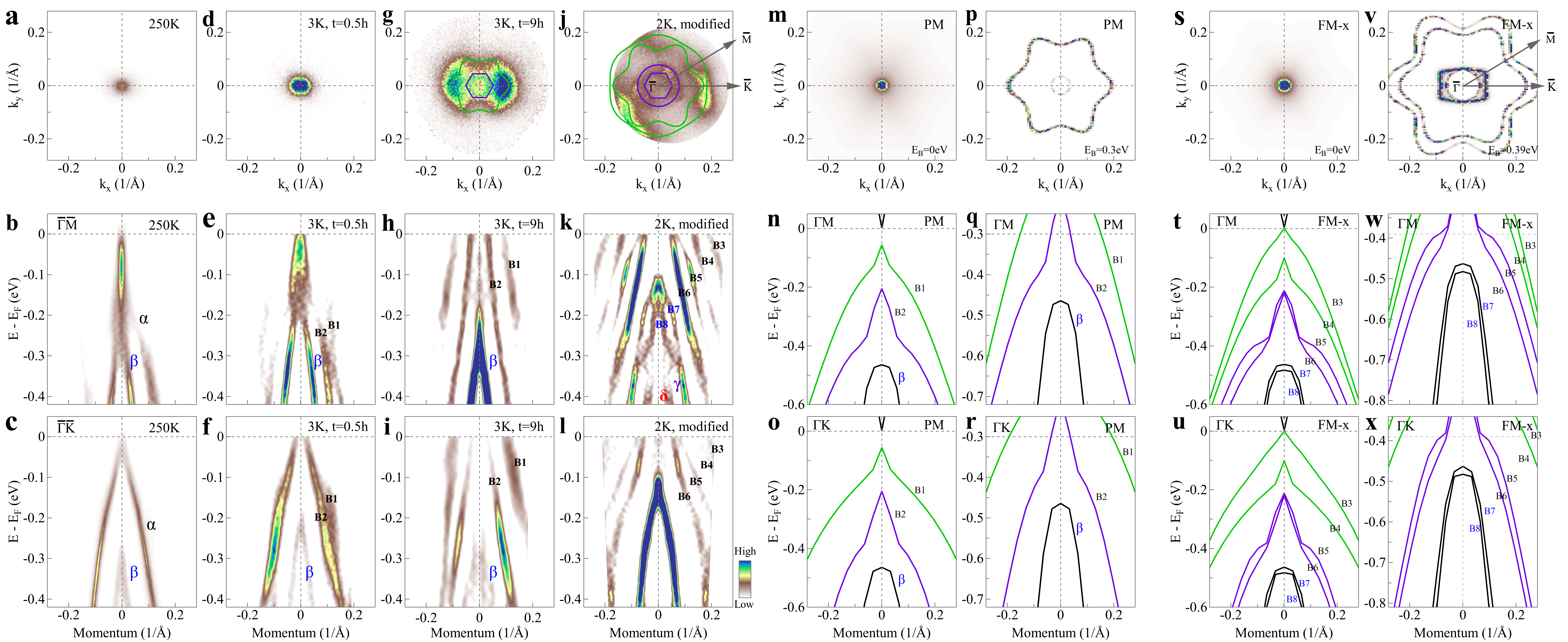}
\end{center}
\caption{\textbf{Comparison of the measured electronic structure of EuCd$_2$As$_2$ with the band structure calculations.} (a) Fermi surface of EuCd$_2$As$_2$ measured at 250\,K (the same data as the first panel of Fig. 1c). The corresponding band structures measured along $\bar\mathrm\Gamma$-$\bar\mathrm{M}$ and $\bar\mathrm\Gamma$-$\bar\mathrm{K}$ directions are shown in (b) and (c), respectively. (d-f) Same as (a-c) but for the sample immediately measured at 3\,K after quickly cooling down from 250\,K (the same data as the second panel of Fig. 1c). (g-i) Same as (a-c) but for the sample that is cooled at 3\,K for 9 hours (the same data as the fifth panel of Fig. 1c). (j-l) Same as (a-c) but for the sample measured at 2\,K after staying for a significantly long time (the same data as the first panel of Fig. 3a). (m-o) Calculated Fermi surface of EuCd$_2$As$_2$ in the paramagnetic state (PM) (m) and the corresponding band structures along $\bar\mathrm\Gamma$-$\bar\mathrm{M}$ (n) and $\bar\mathrm\Gamma$-$\bar\mathrm{K}$ (o) directions. (p) Calculated constant energy contour at the binding energy of 0.3\,eV for EuCd$_2$As$_2$ in the paramagnetic state. The corresponding band structures along $\bar\mathrm\Gamma$-$\bar\mathrm{M}$ and $\bar\mathrm\Gamma$-$\bar\mathrm{K}$ directions at high binding energies of [0.25, 0.72]\,eV are shown in (q) and (r), respectively. (s-u) Same as (m-o) but for EuCd$_2$As$_2$ in the ferromagnetic state with the in-plane magnetization (FM-x) (for the details refer to Fig. S11 and Fig. S12 in Supplementary Materials\cite{SM}). (v) Calculated constant energy contour at the binding energy of 0.39\,eV for EuCd$_2$As$_2$ in the ferromagnetic state with the in-plane magnetization (FM-x). The corresponding band structures along $\bar\mathrm\Gamma$-$\bar\mathrm{M}$ and $\bar\mathrm\Gamma$-$\bar\mathrm{K}$ directions at high binding energies of [0.34, 0.81]\,eV are shown in (w) and (x), respectively.
}
\end{figure*}


\begin{thebibliography}{99}
%%%%%Doping
\bibitem{MImadaRMP} M. Imada, A. Fujimor and Y. Tokura, Metal-insulator transitions, Rev. Mod. Phys. {\bf 70}, 1039 (1998).
\bibitem{PLeeRMP} P. A. Lee, N. Nagaosa and X-G Wen, Doping a Mott insulator: Physics of high-temperature superconductivity, Rev. Mod. Phys. {\bf 78}, 17 (2006).

%%%%% Introduction  EuCd2As2
\bibitem{GYHua_PRB2018} G. Y. Hua, S. Nie, Z. Song, R. Yu, G. Xu and K. Yao, Dirac semimetal in type-IV magnetic space groups, Phys. Rev. B {\bf 98}, 201116(R) (2018).

\bibitem{LLWang_PRB} L. L. Wang, N. H. Jo, B. Kuthanazhi, Y. Wu, R. J. McQueeney, A. Kaminski and P. C. Canfield, Single pair of Weyl fermions in the half-metallic semimetal EuCd$_2$As$_2$, Phys. Rev. B {\bf 99}, 245147 (2019).

\bibitem{CWNiu_PRB} C. Niu, N. Mao, X. Hu, B. Huang and Y. Dai, Quantum anomalous Hall effect and gate-controllable topological phase transition in layered EuCd$_2$As$_2$, Phys. Rev. B {\bf 99}, 235119 (2019).

\bibitem{JZMa_SA} J. Z. Ma, S. M. Nie, C. J. Yi, J. Jandke, T. Shang, M. Y. Yao, M. Naamneh, L. Q. Yan, Y. Sun and A. Chikina \emph{et al}., Spin fluctuation induced Weyl semimetal state in the paramagnetic phase of EuCd$_2$As$_2$, Sci. Adv. {\bf 5}, eaaw4718 (2019).

\bibitem{JZMa_AM} J. Ma, H. Wang, S. Nie, C. Yi, Y. Xu, H. Li, J. Jandke, W. Wulfhekel, Y. Huang and D. West \emph{et al}., Emergence of nontrivial low-energy Dirac Fermions in antiferromagnetic EuCd$_2$As$_2$, Adv. Mater. {\bf 32}, 1907565 (2020).

\bibitem{Soh_PRB2019} J. R. Soh, F. de Juan, M. G. Vergniory, N. B. M. Schr$\ddot{o}$ter, M. C. Rahn, D. Y. Yan, J. Jiang, M. Bristow, P. Reiss and J. N. Blandy \emph{et al}., Ideal Weyl semimetal induced by magnetic exchange, Phys. Rev. B {\bf 100}, 201102(R) (2019).

\bibitem{Artmann_Chem} A. Artmann, A. Mewis, M. Roepke and G. Michels, AM$_2$X$_2$-verbindungen mit CaAl$_2$Si$_2$-struktur. XI. Struktur und eigenschaften der verbindungen ACd$_2$X$_2$ (A: Eu, Yb;  X: P, As, Sb), Z. Anorg. Allg. Chem. {\bf 622}, 679 (1996).

\bibitem{Schellenberg_Chem} I. Schellenberg, U. Pfannenschmidt, M. Eul, C. Schwickert and R. P$\ddot{o}$ttgen, A $^{121}$Sb and $^{151}$Eu M$\ddot{o}$ssbauer spectroscopic investigation of EuCd$_2$X$_2$ (X = P, As, Sb) and YbCd$_2$Sb$_2$, Z. Anorg. Allg. Chem. {\bf 637}, 1863 (2011).

\bibitem{MCRahn_PRB} M. C. Rahn, J.-R. Soh, S. Francoual, L. S. I. Veiga, J. Strempfer, J. Mardegan, D. Y. Yan, Y. F. Guo, Y. G. Shi and A. T. Boothroyd, Coupling of magnetic order and charge transport in the candidate Dirac semimetal EuCd$_2$As$_2$, Phys. Rev. B {\bf 97}, 214422 (2018).

\bibitem{YXu_PRB} Y. Xu, L. Das, J. Z. Ma, C. J. Yi, S. M. Nie, Y. G. Shi, A. Tiwari, S. S. Tsirkin, T. Neupert and M. Medarde \emph{et al}., Unconventional transverse transport above and below the magnetic transition temperature in Weyl semimetal EuCd$_2$As$_2$, Phys. Rev. Lett. {\bf 126}, 076602 (2021).

\bibitem{Jo_PRB} N. H. Jo, B. Kuthanazhi, Y. Wu, E. Timmons, T.-H. Kim, L. Zhou, L.-L. Wang, B. G. Ueland, A. Palasyuk and D. H. Ryan \emph{et al}., Manipulating magnetism in the topological semimetal EuCd$_2$As$_2$, Phys. Rev. B {\bf 101}, 140402(R) (2020).

\bibitem{YWang_CPL2021} Y. Wang, Cong Li, Yong Li, X. Zhou, W. Wu, R. Yu, J. Zhao, C. Yin, Y. Shi and C. Jin \emph{et al}., Long-time magnetic relaxation in antiferromagnetic topological material EuCd$_2$As$_2$, Chin. Phys. Lett. {\bf 38}, 077201 (2021).

\bibitem{Sanjeewa_PRB} L. D. Sanjeewa, J. Xing, K. M. Taddei, D. Parker, R. Custelcean, C. dela Cruz and A. S. Sefat, Evidence of Ba-substitution induced spin-canting in the magnetic Weyl semimetal EuCd$_2$As$_2$, Phys. Rev. B {\bf 102}, 104404 (2020).

\bibitem {SM}See Supplemental Material at [http://link.aps.org/supplemental/10.1103/PhysRevB.106.085134] for detailed electronic structure evolution of EuCd$_2$As$_2$ with the measurement time at 3\,K; identification of the measured Fermi surface for the modified EuCd$_2$As$_2$; electronic structure evolution of EuCd$_2$As$_2$ with the measurement time at 25\,K; band structure evolution of EuCd$_2$As$_2$ with the measurement time at 150\,K; detailed temperature-dependent electronic structure of the modified EuCd$_2$As$_2$; two-dimensional images showing the second-derivative MDCs at different temperatures; detailed analysis of the band structure measured on the modified EuCd$_2$As$_2$ at different temperatures; time-dependent STM measurements of EuCd$_2$As$_2$ at low temperature; time-dependent LEED measurements of EuCd$_2$As$_2$ at low temperature; estimation of the doping level in the modified EuCd$_2$As$_2$; calculated band structure of EuCd$_2$As$_2$ bulk in different magnetic states; calculated Fermi surface and constant energy contours of EuCd$_2$As$_2$ in the ferromagnetic state with an in-plane magnetization; calculated band structure of EuCd$_2$As$_2$ slab in paramagnetic state; calculated band structures of EuCd$_2$As$_2$ monolayer in different magnetic configurations; comparison of the measured and calculated band structures of EuCd$_2$As$_2$ in paramagnetic state.


%%% Discussion
\bibitem{RJiang_PRB2012} R. Jiang, L.-L. Wang, M. Huang, R. S. Dhaka, D. D. Johnson, T. A. Lograsso and A. Kaminski, Reversible tuning of the surface state in a pseudobinary Bi$_2$(Te-Se)$_3$ topological insulator, Phys. Rev. B {\bf 86}, 085112 (2012).


%%%Sample method
\bibitem{HPWang_PRB} H. P. Wang, D. S. Wu, Y. G. Shi and N. L. Wang, Anisotropic transport and optical spectroscopy study on antiferromagnetic triangular lattice EuCd$_2$As$_2$: An interplay between magnetism and charge transport properties, Phys. Rev. B {\bf 94}, 045112 (2016).

%%%ARPES method
\bibitem{GDLiu_RSI2008}G. D. Liu, G. Wang, Y. Zhu, H. Zhang, G. Zhang, X. Wang, Y. Zhou, W. Zhang, H. Liu and L. Zhao \emph{et al}., Development of a vacuum ultra-violet laser-based angle-resolved photoemission system with a super-high energy resolution better than 1 meV, Rev. Sci. Instrum. {\bf 79}, 023105 (2008).

\bibitem{XJZ_Review} X. J. Zhou, S. He, G.-D. Liu, L. Zhao, L. Yu and W. Zhang, New developments in laser-based photoemission spectroscopy and its scientific applications: a key issues review, Rep. Prog. Phys. {\bf 81}, 062101 (2018).

%%% DFT method
\bibitem{GKresse_PRB1996}G. Kresse and J. Furthm$\ddot{u}$ller, Efficient iterative schemes for ab initio total-energy calculations using a plane-wave basis set, Phys. Rev. B {\bf 54}, 11169 (1996).

\bibitem{John_PRL1996}J. P. Perdew, K. Burke and M. Ernzerhof, Generalized Gradient Approximation Made Simple, Phys. Rev. Lett. {\bf 77}, 3865 (1996).

\bibitem{PEBlochl_PRB1994} P. E. Bl$\ddot{o}$chl, Projector augmented-wave method, Phys. Rev. B {\bf 50},17953 (1994).

\bibitem{GKresse_PRB1999} G. Kresse and D. Joubert, From ultrasoft pseudopotentials to the projector augmented-wave method, Phys. Rev. B {\bf 59}, 1758 (1999).

\bibitem{GPizzi_JPCM2020} G. Pizzi, V. Vitale, R. Arita, S. Bl$\ddot{u}$gel, F. Freimuth, G. G$\acute{e}$ranton, Marco Gibertini, D. Gresch, C. Johnson and T. Koretsune \emph{et al}., Wannier90 as a community code: new features and applications, J. Phys. Condens. Matter {\bf 32}, 165902 (2020).

\bibitem{QSWu_CPC2018} Q. S. Wu, S. N. Zhang, H.-F. Song, M. Troyer and A. A. Soluyanov, Wanniertools: An open-source software package for novel topological materials, Comput. Phys. Commun. {\bf 224}, 405 (2018).

\bibitem{HJKulik_PRL2006} H. J. Kulik, M. Cococcioni, D. A. Scherlis and N. Marzari, Density Functional Theory in Transition-Metal Chemistry: A Self-Consistent Hubbard U Approach, Phys. Rev. Lett. {\bf 97}, 103001 (2006).


\end{thebibliography}
\end{document}